\documentclass[12pt,ams]{article}%
\usepackage[nohead]{geometry}
\usepackage{geometry}
\usepackage{dsfont}
\usepackage{amsmath}
\usepackage{amsfonts}
\usepackage{amssymb}
\usepackage{graphicx}%
\usepackage{subfig}
\usepackage{float}
\usepackage{tikz}
\usepackage{epsfig}

\usepackage{physics}

\usepackage{hyperref}
\begin{document}

\date{}
\title{\textbf{Multivalued fields and monopole operators in topological superconductors}}

\author{
\textbf{P. R. Braga}$^{a}$\thanks{pedro.rangel.braga@gmail.com}\,\,,
\textbf{M. S. Guimaraes}$^{a}$\thanks{msguimaraes@uerj.br}\,\,,
\textbf{M. M. A. Paganelly}$^{a}$\thanks{matheuspaganelly@gmail.com}\,\,,
  \\ \small \textnormal{$^{a}$ \it Departamento de F\'{\i }sica Te\'{o}rica, Instituto de F\'{\i }sica, UERJ - Universidade do Estado do Rio de Janeiro} \
   \\ \small \textnormal{\phantom{$^{a}$} \it Rua S\~{a}o Francisco Xavier 524, 20550-013 Maracan\~{a}, Rio de Janeiro, Brasil.}}

\maketitle

 
\begin{abstract}

In this work, we investigate the role of multivalued fields in the formulation of monopole operators and their connection with topological states of matter. In quantum field theory it is known that certain states describe collective modes of the fundamental fields and are created by operators that are often non-local, being defined over lines or higher-dimensional surfaces. For this reason, they may be sensitive to global, topological, properties of the system and depend on nonperturbative data. Such operators are generally known as monopole operators. Sometimes they act as disorder operators because their nonzero expectation values define a disordered vacuum associated with a condensate of the collective modes, also known as defects. In this work we investigate the definition of these operators and their relation to the multivalued properties of the fundamental fields. We study several examples of scalar field theories and generalize the discussion to $p$-forms, with the main purpose of studying new field configurations that may be related to topological states of matter. We specifically investigate the so-called chiral vortex configurations in topological superconductors. We highlight an important aspect of this formalism, which is the splitting of the fields in their regular and singular parts that identifies an ambiguity that can be explored, much like gauge symmetry, in order to define observables.

\end{abstract}

\section{Introduction}

In this work we aim to develop a formalism for dealing with multivalued fields and to study their application in the description of defects in field theory, giving a clearer interpretation of how defects can be used to classify superconductors as topological states of matter.

Quantum field theory can be defined by the algebra of local operators \cite{Haag:1992hx}. Modulo superselection sectors, the theory can also be defined by the correlation functions of local operators, which are usually the quantum versions of the fundamental fields, and composite operators constructed from them, appearing in a classical Lagrangian formulation of the theory. Locality of the Lagrangian formulation demands such operators to be local, defined at a point, but there is important information also in the correlation function of non-local operators, that are defined on extended regions such as lines and surfaces \cite{Mandelstam:1975hb, tHooft:1977nqb, Gukov:2006jk, Gukov:2008sn}. Following current nomenclature, we will generically call them \emph{monopole operators} \cite{Borokhov:2002ib, Borokhov:2002cg}. The name comes from the fact that in compact electrodynamics the operator that creates a monopole state is non-local in the sense of demanding a (Dirac) string for its definition with respect to the original gauge field. It had received other names such as surface operator, disorder operator, vortex operator, etc. In fact, they are akin to disorder operators in lattice Ising systems, where the original variables are the spin operators localized on sites and the disorder operators are defined by a set of links cut by a line. Each cut link changes the coupling between the spins of the sites it connects and it is said that this configuration represents a defect in the system. Since the disorder operator involves a collection of spins in different sites, in terms of the original spin variables it is non-local \cite{Kadanoff:1970kz, Marino:1980rn}. Alternatively, one may devise a local formulation in terms of dual variables in a dual lattice \cite{Savit:1979ny}. This is the usual order-disorder duality. In a quantum field theory language, the local operators create local excitations, particles, and the non-local extended operators create collective excitations of the original particles or localized states in a dual formulation. Monopole operators are tricky to study because the very definition of a fundamental excitation is perturbative but duality exchanges weak and strong couplings. It follows that if the original system is in the perturbative regime, the monopole operators will be local in a formulation that is in a strongly coupled regime. Therefore, these operators are usually studied in their non-local formulation. 

The theory of monopole operators has been developed over the course of many years. Its origins can be traced to the work of Kadanoff and Ceva \cite{Kadanoff:1970kz} in the context of lattice theories. It was subsequently introduced in field theory by Mandelstam \cite{Mandelstam:1975hb} and 't Hooft \cite{tHooft:1977nqb}. The concept was further developed by many authors, of which most notable for our approach were Ezawa \cite{Ezawa:1978kf}, Fr\"ohlich \cite{Frohlich:1987,Frohlich:1982}, Polyakov \cite{Polyakov:1980ca},  Polchinski \cite{Polchinski:1980ig} and Marino and Swieca \cite{Marino:1980rn}. Recently, the subject has received great attention in connection with the Langlands program \cite{Kapustin:2006pk} and with a systematic study of the many uses of monopole operators in gauge theories \cite{Borokhov:2002ib, Borokhov:2002cg, Kapustin:1999ha, Kapustin:2005py, Kapustin:2010pk, Gukov:2013zka}. 

The non-local nature of monopole operators demands an extension of the usual definition of fields. In a Lagrangian setting, it is generally assumed that fields are regular functions without singularities and are infinitely differentiable. However,  Dirac's formulation of monopoles in electromagnetism have shown that it can be useful to allow for multivalued fields. More recently this has been understood as part of a mathematical structure that takes the fields as defined over non-trivial spaces \cite{Freed:2000ta, Bauer:2004nh, Freed:2006yc, Kapustin:2014gua, Gaiotto:2014kfa}, but there are circumstances where allowing the functions to be multivalued may lead to a more direct description of monopoles operators (this is similar to the dichotomy between Dirac's string formulation \cite{Dirac:1948um} and Wu-Yang's formulation \cite{Wu:1976ge} of magnetic monopoles). The study of these instances is the purpose of this work in order to apply it to the newly discovered topological states of matter. 

We will describe multivalued fields with the help of surfaces whose placement in space will characterize the multivalued character of the fields. The formulation presented here is restricted to examples where the field may be expressed as a direct sum of a regular, single-valued, part and a singular, multivalued, part. The different field configurations are thus specified by the continuous field representing the regular part along with the singular part that is characterized by distribution-valued fields that may be identified with surfaces generically called branes. This formulation is not completely new and most of what we do is a recasting of ideas that are already present in \cite{Kadanoff:1970kz, Marino:1980rn, Frohlich:1987, Frohlich:1982}, but the main point is to highlight a symmetry associated with the freedom in defining brane configurations. In the path integral formulation of the system, one regards the ensemble as a mixture of continuous variables, to be integrated over, and branes configurations, to be summed over. In this work we will not try to specify a prescription to evaluate the sum over brane configurations; the subject has been developed by many authors and one interesting recent take on this can be found in \cite{Oxman:2018dzp}. The importance of multivalued fields and their connection with the so-called large gauge symmetry, with the explicit use of singular surfaces along the lines presented here, was pointed out by Kleinert \cite{Kleinert:1990ts, Kleinert:2008zzb}. The concept of singular surfaces and its importance for the structure of the non-perturbative vacuum in gauge theories, taking into account their collective behavior, was discussed in \cite{Banks:1977cc, Quevedo:1996uu} and subsequently further developed with many applications leading to the definition of brane symmetry \cite{Grigorio:2009pi, Grigorio:2011pi, Guimaraes:2012tx, Grigorio:2012jt, Guimaraes:2012ma}. 

The main result of this paper, building on a framework developed by many authors, is the representation of multivalued fields suitable to the analysis of defects that are relevant to the study of topological states of matter like the chiral vortex in topological superconductors. Vortices configurations play a very important role in the formulation of the theory of topological superconductors \cite{ Qi:2012cs}. In particular, one of the main characteristics of a  system with topological superconductivity is the presence of so-called chiral vortices, that do not carry electromagnetic flux \cite{Qi:2012cs, Stone:2016pof, Braga:2016atp}. These chiral vortices decouple at the level of the deep low energy effective action of the system, in the sense that the effective theory for the electromagnetic response of the system is described by a topological action where the relevant degrees of freedom are vortices that carry flux and quasi-particles, as in the normal superconductor \cite{Hansson:2004wca}. Nevertheless, chiral vortices non-trivially trigger the non-conservation of the canonical supercurrent and since the source of this non-conservation resides in a vortex without flux, one can view it as induced by Majorana modes localized on the vortex. Majorana excitations are a hallmark of topological superconductors. This provides an important characterization of this particular superconducting state. In order to better understand the role of these chiral vortices, we present a formalism based on multivalued fields and brane symmetry that clarifies the vortex locus where the chiral configurations appear. 

This work is organized as follows: in section 2, we provide a general definition of what we mean by multivalued fields, Section 3 deals with the concept of brane symmetry and its relation with other symmetries of the system. We will first study the case of the scalar field and its global shift symmetry and then proceed with the study of gauge symmetries. The study of gauge symmetry will naturally lead to the study of brane symmetry as a manifestation of the so-called singular or large gauge transformation.  In section 4, we apply the formalism to the study of superconducting systems. We discuss the important role played by multivalued fields in understanding the electromagnetic response of superconductors and their application in topological superconductivity. In section 5, we present our concluding remarks.

\section{Multivalued fields}

\subsection{Scalar fields}

In this section we study several properties of scalar fields systems when multivalued (singular) contributions are allowed in their definition. We start by considering a scalar field $\phi(x)$ and a closed curve $C$ in such a way that 
\begin{eqnarray}\label{singularity}
\oint_{C}\dd{x^{\mu}}\partial_{\mu}\phi(x)= \alpha,
\end{eqnarray}
where $\alpha$ is a constant. If $\alpha$ is nonzero we see that the field $\phi$ cannot be regular everywhere. One can understand this better using Stokes theorem in order to write
\begin{eqnarray}\label{scalar_field_stokes}
\oint_{C}\dd{x^{\mu}}\partial_{\mu}\phi(x)=\int_{S}\dd{\sigma_{1}}\dd{\sigma_{2}}\varepsilon^{a_{1}a_{2}}\pdv{y^{\mu_{1}}}{\sigma_{a_{1}}}\pdv{y^{\mu_{2}}}{\sigma_{a_{2}}}\partial_{\mu_{1}}\partial_{\mu_{2}}\phi=\alpha,
\end{eqnarray}
where $S$ is the surface bounded by the curve $C$. Thus, if $\alpha \neq 0$, it must be that $\partial_{\left[\mu_{1}\right.}\partial_{\left.\mu_{2}\right]}\phi(x)\neq 0$ at least at some point in $S$. This implies that there is a singular contribution to $\phi(x)$ which can be represented by an additive contribution
\begin{eqnarray}\label{multivalued_scalar_expression}
\phi(x)=\phi^{reg}(x)+\beta \int^{x}\dd{y^{\mu}} \tilde{K}_{\mu}(y; \Upsilon),
\end{eqnarray}
where $\beta$ is a constant and  $\phi^{reg}$ stands for the regular part of the field (meaning $\partial_{\left[\mu_{1}\right.}\partial_{\left.\mu_{2}\right]}\phi^{reg}(x) = 0$). The singular part is represented by the integral over $\tilde{K}$. The expression for $\tilde{K}$ is given by  
\begin{eqnarray}\label{K_tilde_expression}
\tilde{K}_{\mu}(x; \Upsilon)&=&\frac{1}{(D-1)!}\varepsilon_{\mu\mu_{1}\mu_{2}\ldots\mu_{D-1}}K^{\mu_{1}\mu_{2}\ldots\mu_{D-1}}(x; \Upsilon)\nonumber\\
&=&\frac{1}{(D-1)!}\varepsilon_{\mu\mu_{1}\mu_{2}\ldots\mu_{D-1}}\int_{\Upsilon}\dd{\tau_{1}}\ldots\dd{\tau_{(D-1)}\varepsilon^{a_{1}\ldots a_{(D-1)}}}\pdv{y^{\mu_{1}}}{\tau_{a_{1}}}\ldots\pdv{y^{\mu_{D-1}}}{\tau_{a_{(D-1)}}}\delta^{D}(x-y(\tau_{a_{1}},\ldots,\tau_{a_{(D-1)}}))\nonumber\\
&=&\varepsilon_{\mu\mu_{1}\mu_{2}\ldots\mu_{D-1}}\int_{\Upsilon}\dd{\tau_{1}}\ldots\dd{\tau_{(D-1)}}\pdv{y^{\mu_{1}}}{\tau_{1}}\ldots\pdv{y^{\mu_{D-1}}}{\tau_{(D-1)}}\delta^{D}(x-y(\tau_{a_{1}},\ldots,\tau_{a_{(D-1)}})). 
\end{eqnarray}
We see that $\tilde{K}$ is the Hodge dual of $K$, where $K$ is the Poincare dual of the $(D-1)$-surface (codimension-$1$ surface) $\Upsilon$, that is, $K$ is a $(D-1)$-brane. We define a $p$-brane as a singular $p$-form defining a $p$-surface embedded on a $D$-dimensional manifold. 

One can note that the integral defining the singular part of the field in \eqref{multivalued_scalar_expression} is path dependent, that is, we must choose a particular path to integrate over ending on the point in space where we are evaluating the field, and different paths will lead to different results. This defines $\phi$ as a multivalued field. The $(D-1)$-brane $K$ thus defines a codimension-$1$ surface in the $D$-dimensional space such that the intersection number with the line defining the path of integration will provide a nonzero contribution proportional to $\beta$. Explicitly we have
\begin{eqnarray}\label{close_contour}
\oint_{C} dx^{\mu} \partial_{\mu} \phi &=& \beta \oint_{C} dx^{\mu} \tilde{K}_{\mu}\nonumber\\
&=&  \beta \oint_{C} d\tau \int_{\Upsilon} d\tau_1 \cdots d\tau_{D-1} \; \varepsilon_{\mu\mu_1\cdots \mu_{D-1}}  \frac{\partial x^{\mu}}{\partial \tau} \frac{\partial y^{\mu_1}}{\partial \tau_{1}} \cdots \frac{\partial y^{\mu_{D-1}}}{\partial \tau_{D-1}} \delta^D (x(\tau)-y(\tau_1, \cdots, \tau_{D-1})) \nonumber\\
&=& \beta n.
\end{eqnarray}
Where $n \in \mathbb{Z}$ is the number of times that the line $C$ crosses the surface $\Upsilon$. Choosing $\beta = \frac{\alpha}{n}$, we obtain the desired property \eqref{singularity}.  Here we note that the complete information about singularities and the  multivaluedness of the field is encoded in the line integral of $\tilde{K}$. This is due to the fact that the singular part in this case contributes additively to the expression.

The dynamical theory of multivalued fields is constructed imposing brane invariance as a symmetry akin to gauge symmetry. Observables should be invariant under transformations of the brane. Considering a scalar field theory, we modify the field $\phi$ to allow for a multivalued contribution as in \eqref{multivalued_scalar_expression}
\begin{eqnarray}\label{multivalued_scalar_expression2}
\phi(x)=\phi^{reg}(x)+\alpha \int^{x}\dd{y^{\mu}} \tilde{K}_{\mu}(y; \Upsilon),
\end{eqnarray}
Deforming the brane from $\Upsilon$ to $\Upsilon'$,  a surface $\Sigma$ is swept by $\Upsilon$ as it is deformed 
\begin{align}
\label{braneshift}
\int^x dy^{\mu} \tilde{K}_{\mu} (y,\Upsilon)  &\rightarrow  \int^x dy^{\mu} \tilde{K}_{\mu} (y,\Upsilon')  - \Theta(\Sigma, x)
\end{align}
where
\begin{align}
\Theta(x, \Sigma) = \begin{cases} 1 &\mbox{if } x \in \Sigma \\ 
0 & \mbox{if } x \notin \Sigma \end{cases} 
\end{align}
In order for $\phi(x)$ to be an invariant we must have under this deformation also
\begin{align}
\label{scalarmasslessshift2}
\phi^{reg}(x) \rightarrow \phi^{reg}(x) + \alpha\Theta(x, \Sigma) 
\end{align}
Note that $\partial_{\left[\mu \right.} \partial_{\left.\nu \right]} \Theta(\Sigma, x) = 0$ so that regularity of $\phi^{reg}(x)$ is maintained by this transformation. Therefore, under the transformations \eqref{braneshift}, \eqref{scalarmasslessshift2}, $\phi$ defined in  \eqref{multivalued_scalar_expression2} remains invariant, as it should because it is an observable. Brane symmetry is similar to gauge symmetry (in fact, note that the transformation \eqref{scalarmasslessshift2} localizes the global transformation $\phi(x) \rightarrow \phi(x) + \alpha$ over the surface $\Sigma$). Thus, observables must be brane invariant as well as gauge invariant. As we will see shortly, there is an important connection between gauge and brane symmetries.

\subsection{$p$-forms}
We can follow the same lines when discussing general $p$-forms in $D$-dimensions. We define a  multivalued $p$-form as 
\begin{align}
\label{singpform2b}
A_{\mu_1\cdots \mu_p}(x)  = A^{reg}_{\mu_1\cdots \mu_p}(x)  + \frac{\alpha}{(p+1)!} \int^x dy^{\mu} \tilde{K}_{\mu \mu_1\cdots \mu_p} (y; \Sigma)
\end{align}
where
\begin{align}
\label{singpform2ab}
&\tilde{K}_{\mu_1 \mu_2\cdots \mu_{p+1}} (x;\Sigma)
=  \varepsilon_{\mu_1\cdots\mu_{p+1}\nu_{p+2}\cdots \nu_{D}}  \int_{\Sigma} d\tau_1 \cdots d\tau_{D-p-1} \; \frac{\partial y^{\nu_{p+2}}}{\partial \tau_{1}} \cdots \frac{\partial y^{\nu_{D}}}{\partial \tau_{D-p-1}} \delta^D (x-y(\tau_1, \cdots, \tau_{D-p-1}))\nonumber\\
&=   \frac{1 }{(D-p-1)!} \varepsilon_{\mu_1\cdots\mu_{p+1}\nu_{p+2}\cdots \nu_{D}}  \int_{\Sigma} d\tau_1 \cdots d\tau_{D-p-1} \; \frac{\partial y^{\nu_{p+2}}}{\partial \tau_{a_1}} \cdots \frac{\partial y^{\nu_{D}}}{\partial \tau_{a_{D-p-1}}} \varepsilon^{a_1\cdots a_{D-p-1}} \delta^D (x-y(\tau_1, \cdots, \tau_{D-p-1}))\nonumber\\
&=  \frac{1}{(D-p-1)!} \varepsilon_{\mu_1\cdots\mu_{p+1}\nu_{p+2}\cdots \nu_{D}}   K^{\nu_{p+2}\cdots \nu_{D}} (x; \Sigma)
\end{align}
There is a brane symmetry associated with $K$, which is the Poincare dual of the codimension-$p+1$ surface $\Sigma$. There is a natural observable which is the generalized magnetic $(D-p-2)$-current
\begin{align}
\label{magnetic-current}
J^{\mu_1 \mu_2\cdots \mu_{D-p-2}} (x; \partial\Sigma) = \partial_{\mu} K^{\mu \mu_{1}\cdots \mu_{D-p-2}} (x; \Sigma)
\end{align}
such that 
\begin{align}
\label{magnetic-equation}
\varepsilon^{\mu_1\mu_2\cdots\mu_{D-p-2}\mu_{D-p-1} \cdots \mu_{D}}\partial_{\mu_{D-p-1}} \partial_{\mu_{D-p}}   A_{\mu_{D-p+1}\cdots \mu_{D}} = \alpha J^{\mu_1 \mu_2\cdots \mu_{D-p-2}} (x; \partial\Sigma) 
\end{align}
Therefore, the brane symmetry is 
\begin{align}
\label{branesymmetry2}
K^{\mu_{1}\cdots \mu_{D-p-1}} (x; \Sigma) \rightarrow K^{\mu_{1}\cdots \mu_{D-p-1}}(x, \Sigma)   + \partial_{\mu}  M^{\mu \mu_{1}\cdots \nu_{D-p-1}}(x; \Omega)   
\end{align}
in such a way that $\partial\Omega = \Sigma' - \Sigma$. 

Following our considerations from before, if the multivalued $p$-form is to be made invariant under this transformation (this demand is only called for if the $p$-form is an observable, we will see that when the $p$-form is a gauge field only the gauge invariant forms are to be made invariant under brane symmetry as well), we must have 
\begin{align}
\label{branesymmetry3}
\delta_{\Sigma} A^{reg}_{\mu_1\cdots \mu_p}(x)  &= -\alpha \int^x dy^{\mu}  \varepsilon_{\mu \mu_1\cdots\mu_{p}\nu_{p+1}\cdots \nu_{D-1}}   \frac{1}{(D-p-1)!} \partial_{\rho} M^{\rho\nu_{p+1}\cdots \nu_{D-1}}(y;\Omega) \nonumber\\
&= -\alpha P_{\mu_1\cdots \mu_p}(x)
\end{align}
This cannot be written as a simple Heaviside function as before. But we have the properties 
\begin{align}
\label{branesymmetry4}
\partial_{\mu_1}P^{\mu_1\cdots \mu_p}(x) &= 0\nonumber\\
\varepsilon^{\mu_1 \mu_2\cdots\mu_{p}\nu_{p+1}\cdots \nu_{D}}\partial_{\nu_{p+1}}\partial_{\nu_{p+2}}P_{\mu_1\cdots \mu_p}(x) &= 0
\end{align}
where the second property implies that regularity is maintained by the brane transformation.

\subsection{$1$-forms}

Another well known example is the $1$-form gauge field. The analog of  \eqref{scalar_field_stokes} is
\begin{eqnarray}\label{singularity-1form}
\oint_{S}\dd{\sigma_{1}}\dd{\sigma_{2}}\varepsilon^{a_{1}a_{2}}\pdv{y^{\mu_{1}}}{\sigma_{a_{1}}}\pdv{y^{\mu_{2}}}{\sigma_{a_{2}}}F_{\mu_{1} \mu_{2}} = \int_{V}\dd{\tau_{1}}\dd{\tau_{2}}\dd{\tau_{3}}\varepsilon^{a_{1}a_{2}a_{3}}\pdv{y^{\mu_{1}}}{\tau_{a_{1}}}\pdv{y^{\mu_{2}}}{\tau_{a_{2}}}\pdv{y^{\mu_{3}}}{\tau_{a_{3}}}\partial_{\left[ \mu_{1} \right.} F_{\mu_{2} \left.\mu_{3}\right]} = \alpha,
\end{eqnarray}
where $F_{\mu\nu} = \partial_{\left[ \mu \right.}A_{\left.\nu\right]}$ and $\alpha$ is a constant, identified in this case with the magnetic charge.  We recognize the left hand side as the magnetic flux over the closed surface $S$ that encloses the volume $V$. This is just the expression for the existence of a magnetic charge. For this flux to be nonzero we have to allow $A$ to display a singular part. We define the multivalued form
\begin{align}
  \label{sing1form}
A_{\mu}(x)  = A^{reg}_{\mu}(x)  + \frac{\beta}{2} \int^x dy^{\nu} \tilde{M}_{\nu \mu} (y)
  \end{align} 
and therefore
\begin{align}
  \label{sing1form2}
F_{\mu\nu}(x)  = F_{\mu\nu}^{reg}(x)  + \beta \tilde{M}_{\mu \nu} (x),
  \end{align}   
where  we thus recognize $\tilde{M}$ as the Dirac string.

This discussion can of course be conducted at a more modern level recognizing this structure as a fiber bundle with connection and the nonzero flux of the $2$-form $F$ as the second Chern number of the bundle. The relevant mathematical structure to describe what we are about to encounter will naturally lead to concepts such as the Cheeger-Simons group of differential characters and Deligne-Bailinson cohomology classes (see for instance \cite{Freed:2000ta, Bauer:2004nh, Freed:2006yc, Kapustin:2014gua, Gaiotto:2014kfa} and related works), where the singular parts are naturally incorporated by endowing the mathematical spaces with nontrivial topological properties. But it pays off to maintain the discussion in terms of singular parts because it is more convenient for the present purpose as it makes explicit the contribution of points, strings and p-branes that will lead to the construction of the monopole operators. Because of the connection with the original Dirac string, we will call such singular terms as \textit{Dirac branes}.

\section{Symmetries}

\subsection{Turning global symmetry into brane symmetry}

As with gauge symmetry, one can also use global symmetries as a guide to the introduction of brane symmetry. One simply localizes the global symmetry over a surface embedded in space (gauge symmetry would be the localization over a point) and demands the observables to be invariant under this new symmetry. This is straightforward in the case where the symmetry can be cast as a shift in some field. Consider for instance a complex scalar field with global $U(1)$ symmetry
\begin{align}
\label{complexscalar}
\Phi(x) \rightarrow e^{i\alpha}\Phi(x) 
\end{align}
in terms of the phase  $\theta(x)$ of $\Phi (x)= |\Phi|(x) e^{i\theta(x)}$ we obtain
\begin{align}
\label{complexscalar2}
\theta(x) \rightarrow \theta(x) + \alpha
\end{align}
We then turn this global symmetry into a brane symmetry. This amounts to consider a multivalued $\theta$ 
\begin{align}
\label{complexscalar4}
\theta(x) = \theta^{reg}(x) + \alpha \int^x dy^{\mu} \tilde{K}_{\mu} (y,\Upsilon) 
\end{align}
such that the transformation of $\theta^{reg}$ is compensated by the transformation of $K$ (the deformation of the surface $\Upsilon$) 
\begin{align}
\label{complexscalar5}
\theta^{reg}(x) &\rightarrow \theta^{reg}(x) + \alpha\Theta(x, \Sigma) \\
\int^x dy^{\mu} \tilde{K}_{\mu} (y,\Upsilon)  &\rightarrow  \int^x dy^{\mu} \tilde{K}_{\mu} (y,\Upsilon')  - \Theta(\Sigma, x)
\end{align}
where $\partial\Sigma = \Upsilon' - \Upsilon$.
If the original action of the scalar field has the form
\begin{align}
\label{complexscalaraction}
S = \int d^D x \left(| \partial_{\mu} \Phi |^2 + V(\Phi)\right) 
\end{align}
with $\Phi (x) = \Phi^{reg}(x) e^{i\alpha \int^x dy^{\mu} \tilde{K}_{\mu} (y,\Upsilon) }$, we have
\begin{align}
\label{complexscalaraction2}
\partial_{\mu} \Phi  = \left(\partial_{\mu} \Phi^{reg}(x) + i\alpha \tilde{K}_{\mu} (x,\Upsilon) \Phi^{reg}(x)  \right) e^{i\alpha \int^x dy^{\mu} \tilde{K}_{\mu} (y,\Upsilon) }
\end{align}
It follows that the new action, representing the insertion of the disorder operator (or, which is the same, the imposition of brane symmetry) is
\begin{align}
\label{complexscalaraction3}
S \rightarrow \int d^D x \left(| \partial_{\mu} \Phi^{reg}(x) + i\alpha \tilde{K}_{\mu} (x,\Upsilon) \Phi^{reg}(x)  |^2 + V(\Phi) \right) = \int d^D x \left( | \tilde{D}_{\mu} \Phi^{reg}(x) |^2 + V(\Phi) \right)
\end{align}
where
\begin{align}
\label{branecovariantderiv}
\tilde{D}_{\mu} \Phi^{reg}(x) = \partial_{\mu} \Phi^{reg}(x) + i\alpha \tilde{K}_{\mu} (x,\Upsilon) \Phi^{reg}(x) 
\end{align}
is the brane covariant derivative.

This can be taken as the definition of disorder operators in the continuum formulation of the scalar theory. Computing the path integral with the action \eqref{complexscalaraction3}, integrating over $\Phi^{reg}$ and $\Phi^{\ast reg} $ for a given codimension-$1$ $\Upsilon$ brane configuration, one computes correlation functions of operators defined by the codimension-$2$ brane invariant constructed from $\Upsilon$. Geometrically, the brane invariant is the boundary of $\Upsilon$ and generally gives rise to a non-local operator. A special case occurs in $D=2$ dimensions where $\Upsilon$ is a line whose boundary are points, thus defining local disorder operators. In order to compute, for instance, the $2$-point function of the disorder operator, with the insertion points in $x_0$ and $x_0'$ in the complex scalar theory, we just compute the path integral with the above action where $\Upsilon$ is the line connecting $x_0$ and $x_0'$. Due to the brane symmetry, this $2$-point function does not depend on which $\Upsilon$ is actually chosen to perform the computation 
\begin{align}
\label{pathintcomplex}
\langle \mu^\ast (x_0) \mu(x'_0) \rangle = \frac{1}{Z} \int {\cal D} \Phi^{reg}   {\cal D} \Phi^{\ast reg}  e^{- \int d^2 x | \tilde{D}_{\mu} \Phi^{reg}(x) |^2 + V(\Phi) }
\end{align}
In order to compute a general $n$-point function we just choose appropriate $\Upsilon$ ending on the point of insertions (if it is a $1$-point function for instance we take  $\Upsilon$ as the line from infinity to the point of insertion). The complex nature of the operator amounts to endow $\Upsilon$ with an orientation.

\subsection{Gauge invariance and brane symmetry}

These considerations are suitable to be applied to the case when there is a gauge redundancy in the system. Consider an abelian $1$-form gauge theory in $D$-dimensions. It is described by a gauge field $A_{\mu}$ and the theory is such that it is invariant under 
\begin{align}
\label{gaugesymmetry}
A_{\mu} \rightarrow A_{\mu} + \partial_{\mu} \phi 
\end{align}
The field strength $F_{\mu\nu} = \partial_{\mu} A_{\nu} - \partial_{\nu} A_{\mu}$ is the natural invariant constructed from $ A_{\mu}$. But note that in order for $F_{\mu\nu}$ to be an invariant under \eqref{gaugesymmetry}, $\phi$ must be single-valued. If $\phi$ were multivalued we would have    
\begin{align}
\label{gaugesymmetry2}
A_{\mu} \rightarrow A_{\mu} + \partial_{\mu} \phi^{reg} + \alpha \tilde{K}_{\mu}   
\end{align}
and the field strength would change by 
\begin{align}
\label{gaugesymmetryF}
F_{\mu\nu} \rightarrow F_{\mu\nu}  + \alpha \left(\partial_{\mu} \tilde{K}_{\nu}   - \partial_{\nu} \tilde{K}_{\mu}  \right) 
\end{align}
In \eqref{gaugesymmetry2} we have what one may call a singular gauge transformation, which is a combined regular gauge and brane transformations. In order to make sense of this proposal, we have to allow $A_{\mu}$ to be multivalued
\begin{align}
\label{multiA}
A_{\mu}(x)  = A^{reg}_{\mu}(x) + \frac{\alpha}{2} \int^{x} dy^{\nu} \tilde{M}_{\nu\mu}(y)   
\end{align}
This leads to the definition of the gauge and brane invariant
\begin{align}
\label{multiF}
F_{\mu\nu} \equiv F^{reg}_{\mu\nu}  + \alpha \tilde{M}_{\mu\nu}
\end{align}
Now, eq.\eqref{gaugesymmetryF} becomes the transformation for $ F^{reg}_{\mu\nu}$ and the full set of brane symmetries is
\begin{align}
\label{branesymmetryAF}
\tilde{M}_{\mu\nu} &\rightarrow \tilde{M}_{\mu\nu} - \left( \partial_{\mu} \tilde{K}_{\nu}   - \partial_{\nu} \tilde{K}_{\mu} \right)\\
F^{reg}_{\mu\nu} &\rightarrow F^{reg}_{\mu\nu}  + \alpha \left(\partial_{\mu} \tilde{K}_{\nu}   - \partial_{\nu} \tilde{K}_{\mu}  \right) 
\end{align}
Also, eq.\eqref{gaugesymmetry2} becomes the transformation for $A^{reg}_{\mu}$
\begin{align}
\label{branesymmetryAF2}
A^{reg}_{\mu} &\rightarrow A^{reg}_{\mu} +\partial_{\mu} \phi^{reg} + \alpha \tilde{K}_{\mu} 
\end{align}
while $A_{\mu}$ in eq.\eqref{multiA} transforms as  
\begin{align}
\label{branesymmetryAF3}
A_{\mu} &\rightarrow A_{\mu} +\partial_{\mu} \phi^{reg} + \alpha \tilde{K}_{\mu}  - \alpha \tilde{P}_{\mu}    
\end{align}
where
\begin{align}
\label{branesymmetryAF4}
\tilde{P}_{\mu}   = \frac{1}{2} \int^{x} dy^{\nu} \partial_{\left[ \nu \right.}\tilde{K}_{\left.\mu \right] }(y)   
\end{align}
Note that this in fact maintains $F_{\mu\nu}$ on eq.\eqref{multiF}, invariant, as it must by construction.

It is important to note that this set of transformations is different from the ones discussed in eqs.\eqref{branesymmetry3} and \eqref{branesymmetry4}. There we had a pure brane symmetry that left $A_{\mu}$ invariant, because it was an observable in that case. The main difference is that the pure brane symmetry did not shift the longitudinal part, only the transverse part. In fact, for the present case we would have simply
\begin{align}
\label{branesymmetryAF5}
A^{reg}_{\mu} & \xrightarrow{brane} A^{reg}_{\mu} - \alpha P_{\mu}  
\end{align}
Note that $\partial_{\mu}P^{\mu} =0$, and therefore this is a transformation that shifts only the transverse part of  $A^{reg}_{\mu}$, as opposed to a regular gauge transformation that shifts only the longitudinal part. But when we consider a singular gauge transformation, we include also a transverse shift because of the presence of the extra term $\tilde{K}_{\mu}$ in the expression of $\partial_{\mu}\phi$. The appearance of $P_{\mu}$ has the role of restoring the pure longitudinal shift that a gauge symmetry is supposed to have, in fact
\begin{align}
\label{branesymmetryAF6}
\partial_{\left[ \nu \right.}\tilde{K}_{\left.\mu \right] } - \partial_{\left[ \nu \right.}\tilde{P}_{\left.\mu \right] } = 0
\end{align}
 so that the transverse part suffers no shift under the singular gauge transformation, which can also be seen as a combination of a regular gauge transformation (shift by $\partial_{\mu} \phi^{reg}$) and a brane transformation that only shifts the longitudinal part (shift by $\alpha\tilde{K}_{\mu} -\alpha \tilde{P}_{\mu}$).

Summarizing, we have just constructed the elements to describe the theory of magnetic monopoles in an abelian gauge theory. We learned that allowing for multivalued gauge transformations (multivalued $\phi$) and imposing that the theory remains invariant we are led to introduce multivalued gauge fields. This leads to the definition of the (brane and gauge) invariant $F_{\mu\nu}$ \eqref{multiF}. We thus obtain the identity  
\begin{align}
\label{multiFmono}
\partial_{\mu_1}   \tilde{F}^{\mu_1\mu_2\cdots \mu_{D-2}} = \alpha \partial_{\mu_1} M^{\mu_1\mu_2\cdots \mu_{D-2}} = \alpha J_{m}^{\mu_2\cdots \mu_{D-2}}
\end{align}
Where $J_{m}^{\mu_2\cdots \mu_{D-2}} = \partial_{\mu_1} M^{\mu_1\mu_2\cdots \mu_{D-2}} $ is the magnetic $(D-3)$-current with $M^{\mu_1\mu_2\cdots \mu_{D-2}} $  the Dirac $(D-2)$-brane. In $D=4$ we have the familiar equation 
\begin{align}
\label{multiFmono2}
\partial_{\mu} \tilde{F}^{\mu\nu} = \alpha \partial_{\mu} M^{\mu\nu} = \alpha J_{m}^{\nu}
\end{align}
Where $J_{m}^{\nu} = \partial_{\mu} M^{\mu\nu}$ is the magnetic current with $M^{\mu\nu}$  the world volume of the Dirac string.

\subsubsection{Turning global symmetry into a singular gauge symmetry}

Let us return to our example of the complex scalar field with $U(1)$ global symmetry acting as  
\begin{align}
\label{complexscalara}
\Phi(x) \rightarrow e^{i\alpha}\Phi(x) 
\end{align}
or, in terms of the phase  $\theta(x)$ of $\Phi (x)= |\Phi|(x) e^{i\theta(x)}$ 
\begin{align}
\label{complexscalar2a}
\theta(x) \rightarrow \theta(x) + \alpha
\end{align}
If we now turn this into a local symmetry $\alpha \rightarrow \alpha\phi(x)$ we need to introduce a gauge field and redefine the derivative to a covariant one 
\begin{align}
\label{gaugecovariantderiv}
D_{\mu} \Phi(x) = \partial_{\mu} \Phi(x) - i\alpha A_{\mu} (x) \Phi(x) 
\end{align}
such that the gauge symmetry manifests itself as 
\begin{align}
\label{gaugesym}
\theta(x) &\rightarrow \theta(x) + \alpha\phi(x)\nonumber\\
A_{\mu} &\rightarrow A_{\mu} + \partial_{\mu}  \phi
\end{align}
so that the covariant derivative transforms as a normal derivative would if the symmetry was global
\begin{align}
\label{gaugetransfcovariantderiv}
D_{\mu} \Phi(x) \rightarrow e^{i\alpha\phi(x)} D_{\mu} \Phi(x) 
\end{align}

Now, what if $\phi(x)$ is a multivalued field? This would correspond to a so-called \emph{large gauge transformation}. When we do not consider the multivalued nature of $\phi$ we are in fact regarding it as an infinitesimal parameter and in this case $\phi$ takes value in $\mathbb{R}$, so that this is a $\mathbb{R}$ gauge symmetry. But $\phi$ has the nature of an angular variable since it is a shift  in a field that takes values in the circle $S^1$. So in order to have a true $U(1)$ gauge transformation we must allow $\phi$ to take values in $S^1$, thus defining a $U(1)$ gauge symmetry. This case is also called compact gauge symmetry (because $\phi$ takes values in a compact space) and the former case is called non-compact gauge symmetry. We therefore write
\begin{align}
\label{largegauge}
\phi(x) = \phi^{reg}(x) + \beta\int^x dy^{\mu} \tilde{K}_{\mu} (y)
\end{align}
with $\beta$ a new parameter. The previous analysis of the singular gauge symmetry follows. We already learned that to make sense of a singular gauge transformation we have to introduce a multivalued gauge field
\begin{align}
\label{multiA2}
A_{\mu}(x)  = A^{reg}_{\mu}(x) + \frac{\beta}{2} \int^{x} dy^{\nu} \tilde{M}_{\nu\mu}(y)   
\end{align}
The singular gauge transformation then manifests itself as 
\begin{align}
\label{largegauge2}
\theta(x) &\rightarrow \theta(x) + \alpha\phi^{reg}(x) + \alpha\beta\int^x dy^{\mu} \tilde{K}_{\mu} (y)\nonumber\\
\tilde{M}_{\mu\nu} &\rightarrow \tilde{M}_{\mu\nu} - \left( \partial_{\mu} \tilde{K}_{\nu}   - \partial_{\nu} \tilde{K}_{\mu} \right)\nonumber\\
A^{reg}_{\mu} &\rightarrow A^{reg}_{\mu} +\partial_{\mu} \phi^{reg} + \beta \tilde{K}_{\mu}  
\end{align}
In order to have the expected property, the covariant derivative must be defined with respect to the regular part of the gauge field
\begin{align}
\label{gaugecovariantderiv2}
D_{\mu} \Phi(x) &= \partial_{\mu} \Phi(x) - i\alpha A^{reg}_{\mu} (x) \Phi(x) 
\end{align}
This follows from observing that the relevant invariant structure is
\begin{align}
\label{gaugecovariantderiv3}
\partial_{\mu} \theta(x) - \alpha A^{reg}_{\mu} 
\end{align}
It is thus immediate to see that under eq.\eqref{largegauge2} 
\begin{align}
\label{largegaugetransfcovariantderiv}
D_{\mu} \Phi(x) \rightarrow e^{i\alpha\phi(x)} D_{\mu} \Phi(x) 
\end{align}
It is useful to write the covariant derivative separating the regular parts from the singular ones 
\begin{align}
\label{gaugecovariantderiv4}
D_{\mu} \Phi(x) = e^{i\gamma \int^x dy^{\mu} \tilde{R}_{\mu} (y) }\left( \partial_{\mu} \Phi^{reg}(x) +i\gamma \tilde{R}_{\mu}(x)\Phi^{reg}(x)  - i\alpha A^{reg}_{\mu} (x) \Phi^{reg}(x) \right)
\end{align}
where we defined $\Phi (x) = \Phi^{reg}(x) e^{i\gamma \int^x dy^{\mu} \tilde{R}_{\mu} (y) }$, which amounts to make explicit the singular part of the phase field 
\begin{align}
\label{singphase}
\theta(x) = \theta^{reg}(x) + \gamma\int^x dy^{\mu} \tilde{R}_{\mu} (y)
\end{align}
This is such that under the singular gauge transformation we have
\begin{align}
\label{largegauge3}
\theta^{reg}(x) &\rightarrow \theta^{reg}(x) + \alpha\phi^{reg}(x)\nonumber\\
 \tilde{R}_{\mu} &\rightarrow   \tilde{R}_{\mu}  + \frac{\alpha\beta}{\gamma}\tilde{K}_{\mu} 
\end{align}
Note that we defined a seemingly independent parameter $\gamma$ for the singular part of $\theta$, but in fact it does not make sense to have $\gamma =0$ and $\beta \neq 0$, because even if we start with a regular $\theta$ it will become singular under the multivalued gauge transformation. Therefore  $\gamma$ is not independent from $\beta$ and we can simply identify them $\gamma = \beta$

An important property to note is that the combination $B_{\mu} \equiv A^{reg}_{\mu} (x)- \frac{\gamma}{\alpha} \tilde{R}_{\mu}(x)$ transforms as a regular gauge field under the singular gauge transformation
\begin{align}
\label{largegauge4}
B_{\mu} &\rightarrow B_{\mu} +\partial_{\mu} \phi^{reg}
\end{align}
and the singular gauge covariant derivative assumes the form
\begin{align}
\label{gaugecovariantderiv5}
D_{\mu} \Phi(x) = e^{i\gamma \int^x dy^{\mu} \tilde{R}_{\mu} (y) }\left( \partial_{\mu} \Phi^{reg}(x) - i\alpha B_{\mu} (x) \Phi^{reg}(x) \right)
\end{align}
Such that all the singular character of this derivative is isolated in an overall phase that does not contribute to the action. This is an important property because when computing quantum corrections generated by fluctuations of  $\Phi$, the result will be as if there where no singular parts and the gauge field is $B_{\mu}$.
We can also write the covariant derivative as 
\begin{align}
\label{gaugecovariantderiv6}
D_{\mu} \Phi(x) = e^{i\gamma \int^x dy^{\mu} \tilde{R}_{\mu} (y) }\left( e^{i\theta^{reg}}\partial_{\mu} |\Phi^{reg}(x)| + \left(i\partial_{\mu}\theta^{reg}  - i\alpha B_{\mu} (x) \right) \Phi^{reg}(x) \right)
\end{align}

We can identify three layers of structure for the $U(1)$ scalar field theory:
\begin{enumerate}
\item \emph{Configurations with small $\theta(x)$ and also small gauge transformations $\phi(x)$.} This corresponds to the textbook case where only small fluctuations of $\theta(x)$ are relevant and the gauge group is effectively $\mathbb{R}$, also known as non-compact gauge theory. The analysis can proceed with trivial perturbation theory with all fields being regular. The key quantities to analyze are the gauge invariant terms. Consider the covariant derivative, for instance. The relevant quantity is 
 \begin{align}
 \label{gaugecovariantderiv7}
 \partial_{\mu}\theta^{reg}  - \alpha A^{reg}_{\mu} (x)+ \beta \tilde{R}_{\mu}(x) 
 \end{align} 
with $\theta(x)$ small, we have $\tilde{R}_{\mu}(x) = 0$ and this is just a term gauge equivalent to $A^{reg}_{\mu} (x)$. Also, since the gauge transformation is small, the gauge field is single-valued and $A_{\mu} = A^{reg}_{\mu}$ so that $F_{\mu\nu}(A) =  F_{\mu\nu}(A^{reg})$ and (shifting $ A^{reg}_{\mu} \rightarrow  A_{\mu}^{reg} + \frac{1}{\alpha}\partial_{\mu}\theta^{reg}$)
\begin{align}
\label{fieldstrenght}
F_{\mu\nu}(A^{reg} + \frac{1}{\alpha}\partial_{\mu}\theta^{reg}) = F_{\mu\nu}(A^{reg}) 
\end{align} 
and no nontrivial fluxes are present.

\item \emph{Configurations with large $\theta(x)$ and small gauge transformations $\phi(x)$.} In this case the field $\theta$ takes values in the circle $S^1$ (its image is defined in $[0,2\pi)$), but we only consider infinitesimal gauge transformations. This corresponds to the case where there are vortices in the system associated with closed flux lines. Since $\theta(x)$  is multivalued we write 
\begin{align}
\label{singphasea}
\theta(x) = \theta^{reg}(x) + \beta\int^x dy^{\mu} \tilde{R}_{\mu} (y)
\end{align}
and there is a brane symmetry under which $\theta$ is invariant
\begin{align}
\label{largegauge3a}
\theta^{reg}(x) &\rightarrow \theta^{reg}(x) + \beta\Theta(x)\nonumber\\
 \tilde{R}_{\mu} &\rightarrow   \tilde{R}_{\mu}  - \partial_{\mu}\Theta(x)
\end{align}
This transformation has nothing to do with the gauge symmetry that remains the trivial one $\phi = \phi^{reg}$ and there is no need to consider a multivalued $A_{\mu}$, so that $A_{\mu} = A^{reg}_{\mu}$ (note that the gauge group is still effectively $\mathbb{R}$). But now with nonzero $\tilde{R}_{\mu}$, we have nontrivial fluxes. In fact, with $ B_{\mu}(x)  = A^{reg}_{\mu} (x)   - \frac{\beta}{\alpha} \tilde{R}_{\mu}(x)$
\begin{align}
\label{fieldstrenght2}
F_{\mu\nu}(A) = F_{\mu\nu}(B + \frac{\beta}{\alpha} \tilde{R} ) 
\end{align} 
And the fluxes are closed because we still have
\begin{align}
\label{fieldstrenght3}
\partial_{\mu}  {}^{\ast}F^{\mu\nu}(A) = 0
\end{align}

\item \emph{Configurations with large $\theta(x)$ and large gauge transformations $\phi(x)$.} In this case, both $\theta(x)$ and $\phi(x)$ are compact variables and now both functions are multivalued. $\phi(x)$ defines a compact gauge group $U(1)$. The multivalued gauge transformation demands that $A_{\mu}$ is also multivalued and we have the following set of brane-gauge fields
\begin{align}
\label{largegaugefields}
\theta(x) &= \theta^{reg}(x) + \beta\int^x dy^{\mu} \tilde{R}_{\mu} (y)\nonumber\\
A_{\mu}(x)  &= A^{reg}_{\mu}(x) + \frac{\beta}{2} \int^{x} dy^{\nu} \tilde{M}_{\nu\mu}(y)   
\end{align}
The multivalued gauge transformation
\begin{align}
\label{largegaugefields2}
\phi(x) &= \phi^{reg}(x) + \beta\int^x dy^{\mu} \tilde{K}_{\mu} (y)
\end{align}
acts on these fields as
\begin{align}
\label{largegaugefieldstransf}
\theta^{reg}(x) &\rightarrow \theta^{reg}(x) + \alpha\phi^{reg}(x)\nonumber\\
 \tilde{R}_{\mu} &\rightarrow   \tilde{R}_{\mu}  + \alpha\tilde{K}_{\mu}\nonumber\\
 A^{reg}_{\mu} &\rightarrow A^{reg}_{\mu} +\partial_{\mu} \phi^{reg} + \beta \tilde{K}_{\mu} \nonumber\\   
\tilde{M}_{\mu\nu} &\rightarrow \tilde{M}_{\mu\nu} - \left( \partial_{\mu} \tilde{K}_{\nu}   - \partial_{\nu} \tilde{K}_{\mu} \right)
\end{align}
The covariant derivative remains the same. The term written in \eqref{gaugecovariantderiv7} is invariant under the multivalued gauge transformation. But now we have $A_{\mu} \neq  A_{\mu}^{reg}$ and the field strength has nontrivial open fluxes 
\begin{align}
\label{fieldstrenght4}
F_{\mu\nu}(A) = F_{\mu\nu}(A^{reg}) + \beta   \tilde{M}_{\mu\nu}
\end{align} 
Writing this in terms of the brane invariant gauge field 
\begin{align}
 \label{gaugecovariantderiv7b}
  B_{\mu}(x)  = A^{reg}_{\mu} (x)   - \frac{\beta}{\alpha} \tilde{R}_{\mu}(x) 
 \end{align} 
 we have 
\begin{align}
\label{fieldstrenght5}
F_{\mu\nu}(A) = F_{\mu\nu}(B  + \frac{\beta}{\alpha} \tilde{R} ) + \beta   \tilde{M}_{\mu\nu} =  F_{\mu\nu}(B) + \beta \left( \tilde{M}_{\mu\nu} + \frac{1}{\alpha} \partial_{\left[ \nu \right.}\tilde{R}_{\left.\mu \right] } \right)
\end{align} 
The last term represents brane invariant flux open lines. That they are open can be seen by noting 
\begin{align}
\label{openfluxes}
\partial_{\mu} ^{\ast}F^{\mu\nu}(A) = \beta \partial_{\mu} M^{\mu\nu} = \beta J_{m}^{\nu}
\end{align}

\end{enumerate}

\section{Defects structure in superconductors}
An interesting application of our discussion is the physics of the electromagnetic response in a  superconductor. The effective theory in the London limit is 
\begin{align}
\label{SC}
     S_{SC} &=\int d^{4}x\bigg(\frac{1}{4}F_{\mu\nu}F^{\mu\nu}+\frac{q^2M^{2}}{2}\bigg(A_{\mu}+\frac{1}{q}\partial_{\mu}\theta\bigg)^{2}\bigg) \; ,
\end{align}
This theory is obtained as a limit of the Ginzburg-Landau model such that the field $\theta$ is the phase of the complex scalar field effectively describing the condensate of Cooper pairs. The discussion in the previous section follows almost immediately then. Including vortex contribution to this action amounts to allow $\theta$ to be multivalued leading to closed fluxes contributions as we saw in the item $2$ above. We can also open these flux lines by introducing monopoles contribution making the gauge transformation multivalued as in \eqref{largegaugefields2}, this corresponds to item $3$ above. The action is changed to have the form
\begin{align}
\label{SClg}
     S^{Mon}_{SC} &=\int d^{4}x\bigg(\frac{1}{4}\left( F_{\mu\nu}(B) + \beta \left( \tilde{M}_{\mu\nu} + \frac{1}{\alpha} \partial_{\left[ \nu \right.}\tilde{R}_{\left.\mu \right] } \right)\right)^2 +\frac{q^2M^{2}}{2} \bigg(B_{\mu}+\frac{1}{q}\partial_{\mu}\theta^{reg}\bigg)^2\bigg) \; ,
\end{align}
The last term is a mass term and we see that $\theta^{reg}$ is a Goldstone mode setting the longitudinal part of the gauge field exactly to zero inside the superconductor, while the transverse part decays exponentially within a distance $\sim 1/M$. In fact, variation of the action \eqref{SClg} with respect to changes in $\theta^{reg}$ leads to 
\begin{eqnarray}\label{eqmot-theta}
qM^2 \partial^{\mu}  \bigg(B_{\mu}+\frac{1}{q}\partial_{\mu}\theta^{reg}\bigg)  = 0
\end{eqnarray}
which is the expression for the conservation of the supercurrent  
\begin{eqnarray}\label{supcurrent}
j_s^{\mu} = qM^2  \bigg(B_{\mu}+\frac{1}{q}\partial_{\mu}\theta^{reg}\bigg) 
\end{eqnarray}

The action \eqref{SClg} describes magnetic monopoles inside a superconductor and it is suitable for the computation of the correlation function of monopole operators. In the static case (effectively $D=3$),  $\tilde{M}$ is a line and we can compute the $2$-point function of insertion points connected by this line
\begin{align}
    \langle \mu(x_{0}) \mu(x'_{0}) \rangle = \frac{1}{Z} \sum_{\tilde{R}} \int \mathcal{D}B \; e^{-S^{Mon}_{SC}} 
\end{align} 
In this case $\mu(x_{0})$ denotes the insertion of a monopole at the position $x_0$. $\tilde{M}$ is the Dirac string connecting the monopoles and the sum over $\tilde{R}$ span the closed flux lines configurations and will effectively amount to a sum over all possible lines connecting the monopoles. This correlation function is well known and it is just the 't Hooft loop for confined monopoles; it will have an area law asymptotically. A nice way to see this is to perform a dual transformation obtaining the massive Kalb-Ramond field theory minimally coupled to the line $ \tilde{M}_{\mu\nu} + \frac{1}{\alpha} \partial_{\left[ \nu \right.}\tilde{R}_{\left.\mu \right] } $, thus showing that the line carries energy and its preferred configuration will be the line minimizing this energy, corresponding to a confining string between monopoles.

It is tempting to consider $\mu(x_{0})$ as the disorder operator for a superconductor, but this is misleading because as argued in \cite{Hansson:2004wca}  there is no proper local order parameter for superconductors and the system is better described as a topological state of matter. The main argument of  \cite{Hansson:2004wca} is that the would be order parameter, the complex field in the Ginzburg-Landau model, is gauge dependent. In the present picture the would be disorder field $\mu(x_{0})$ is also nonlocal, since it comes with a choice of the line $\tilde{M}$. In the non-superconducting phase, the Dirac string  $\tilde{M}$ carries no energy and is a trivial redundancy (what we called brane symmetry above), with the closed vortices $\tilde{R}$ decoupling (as can be easily seen with $M=0$ in\eqref{SClg} and rewriting the action in terms of $A^{reg}$). In the superconducting phase the Dirac string  $\tilde{M}$ combines with the closed vortices $\tilde{R}$ to form line configurations that carries energy, but still maintaining brane symmetry, we may call this situation \textsl{brane symmetry breaking}, due to its similarity with the gauge symmetry breaking. But, as in the Higgs mechanism, of course there is no true symmetry breaking and the brane symmetry is just hidden. Gauge symmetry is more properly understood as a redundancy in the variables describing the system and so is the brane symmetry, both are never truly broken. As a result there is no true local disorder parameter.

\subsubsection{Topological superconductor}

An effective theory describing the electromagnetic response of topological superconductor was proposed in \cite{Qi:2012cs}  and further studied in \cite{Stone:2016pof} and \cite{Braga:2016atp}. Here we follow the notation of \cite{Braga:2016atp} where this action was obtained from symmetry considerations and a careful analysis of the relevant degrees of freedom
\begin{align} 
\label{QWZ}
     S_{TSC} &=\int d^{4}x\bigg(\frac{1}{4}F_{\mu\nu}F^{\mu\nu}+\frac{q^2M^{2}}{2}\bigg(A_{\mu}+\frac{1}{q}\partial_{\mu}\theta\bigg)^{2}+\frac{m^{2}}{2}\big(\partial_{\mu}\overline{\theta}\big)^{2}\nonumber\\
     & -i\frac{q^2 M^{4}m^{2}}{\Lambda^{6}}\partial_{\mu}\overline{\theta}\varepsilon^{\mu\nu\rho\sigma}\bigg(A_{\nu}+\frac{1}{q}\partial_{\nu}\theta\bigg)\partial_{\rho}\bigg(A_{\sigma}+\frac{1}{q}\partial_{\sigma}\theta\bigg)+\tilde{\rho}\cos{(\overline{\theta})}\bigg) \; ,
\end{align}
In \eqref{QWZ} we see that the first two terms characterize the usual electromagnetic response of a  superconductor in the London limit, with the real field $\theta$ as the phase of a complex scalar field in the Ginzburg-Landau theory. The other terms characterize the system as a topological superconductor associated with two geometrically disconnected Fermi surfaces (two Fermi surfaces for short). The field $\bar{\theta}$ describes the phase difference between these surfaces and is associated to a charge exchange induced by instantons (see \cite{Braga:2016atp} for details). The massive parameter $M$ quantifies the inverse of the penetration length characterizing the Meissner effect. The Axion-like $\bar{\theta}$ excitations are massive with a mass given by $\sqrt{\frac{\tilde{\rho}}{m^2}}$. The scale $\Lambda$ in the axionic interaction turns out to be not independent, for topological reasons, and given by an integer multiple of $\left(8\pi M^4m^2\right)^{1/6}$. The last term is a Josephson term.

The same analysis including multivalued fields can be carried out for this model. The inclusion of closed vortices $\tilde{R}$, turning $\theta$ into a multivalued field, is straightforward and the introduction of monopoles, opening the flux lines, is also simple, replacing the field $A$ by $B$ everywhere. The result is
\begin{align} 
\label{QWZ2}
     S^{Mon}_{TSC} &=\int d^{4}x\bigg(\frac{1}{4}\left( F_{\mu\nu}(B) + \beta \left( \tilde{M}_{\mu\nu} + \frac{1}{\alpha} \partial_{\left[ \nu \right.}\tilde{R}_{\left.\mu \right] } \right)\right)^2 +\frac{q^2M^{2}}{2} \bigg(B_{\mu}+\frac{1}{q}\partial_{\mu}\theta^{reg}\bigg)^2\nonumber\\
     &+\frac{m^{2}}{2}\big(\partial_{\mu}\overline{\theta}\big)^{2} -i\frac{q^2 M^{4}m^{2}}{\Lambda^{6}}\partial_{\mu}\overline{\theta}\varepsilon^{\mu\nu\rho\sigma}\bigg(B_{\nu}+\frac{1}{q}\partial_{\nu}\theta^{reg}\bigg)\partial_{\rho}\bigg(B_{\sigma}+\frac{1}{q}\partial_{\sigma}\theta^{reg}\bigg)+\tilde{\rho}\cos{(\overline{\theta})}\bigg) \; ,
\end{align} 
But now there is also the possibility of vortices contributions coming from $\overline{\theta}$. These are called \textsl{chiral vortices}. There is no corresponding monopoles for these vortices lines since they do not carry any flux. But they nevertheless have an important contribution to the physics of such superconductors. Computing the variation of the action \eqref{QWZ2} with respect to changes in $\theta^{reg}$ leads to
\begin{eqnarray}\label{eqmot-theta2}
qM^2 \partial^{\mu}  \bigg(B_{\mu}+\frac{1}{q}\partial_{\mu}\theta^{reg}\bigg)  + i\frac{q M^{4}m^{2}}{\Lambda^{6}} \partial_{\mu} \partial_{\nu}\overline{\theta}\;\varepsilon^{\mu\nu\rho\sigma} \partial_{\rho} B_{\sigma} = 0
\end{eqnarray}
We note that if $\overline{\theta}$ is regular, this becomes simply the expression for the conservation of the supercurrent $j_s^{\mu} = qM^2   \bigg(B^{\mu}+\frac{1}{q}\partial^{\mu}\theta^{reg}\bigg) $ of a normal superconductor \eqref{eqmot-theta}.  
Now, consider that $\overline{\theta}$ is multivalued. In that case we have
\begin{align}
\label{chiralvortex}
   \overline{\theta}(x)&=\overline{\theta}^{reg}(x)+\lambda \int^{x}dy^{\mu}\tilde{N}_{\mu}(y) \; ,
\end{align} 
and \eqref{eqmot-theta2} becomes
\begin{align}
\label{QWZ5}
   \partial_{\nu} j_{s}^{\nu} = -  i\frac{q M^{4}m^{2}}{\Lambda^{6}} \lambda  \partial_{\nu}\tilde{N}_{\mu}\varepsilon^{\mu\nu\rho\sigma}\partial_{\rho}B_{\sigma}  =   i\frac{q M^{4}m^{2}}{2\Lambda^{6}} \lambda  \bar{J}_v^{\rho\sigma}F_{\rho\sigma}(B)\; ,
\end{align}
where
\begin{eqnarray}\label{N_tilde_expression}
\tilde{N}_{\mu}(x)&=&\frac{1}{3!}\varepsilon_{\mu\nu\rho\sigma}N^{\nu\rho\sigma}(x)\nonumber\\
&=&\varepsilon_{\mu\nu\rho\sigma} \int_{\Sigma}\dd{\tau_{1}}\dd{\tau_{2}}\dd{\tau_{3}}\pdv{y^{\nu}}{\tau_{1}}\pdv{y^{\rho}}{\tau_{2}}\pdv{y^{\sigma}}{\tau_{3}}\delta^{4}(x-y(\tau_{1}, \tau_{2}, \tau_{3}))
\end{eqnarray}
and the chiral vortex current is defined by
\begin{eqnarray}\label{Jv_expression}
\bar{J}_v^{\mu\nu} = \partial_{\rho} N^{\rho\mu\nu} = \varepsilon^{\mu\nu\rho\sigma} \partial_{\rho}\tilde{N}_{\sigma}
\end{eqnarray}
We thus see that the longitudinal part is not exactly zero inside a topological superconductor, but lives only at vortices loci.

\section{Conclusion}

In this work we have explored the application of multivalued fields in the formulation of monopole operators in the continuum. These operators have a non-local structure that encapsulates collective properties of the original degrees of freedom. The main element in our construction was the introduction of defects through the representation of the fundamental fields as the direct sum of a regular part and a singular part that characterizes their multivaluedness. To the singular part one can associate a geometrical picture in terms of surfaces, or branes, whose arbitrary placement in space is the embodiment of the multivalued nature of the field. It is thus possible to identify different types of ambiguities in the placement of the brane related to the nature of the field: If the field is an observable, that is, if it creates a physical state, there must be no ambiguity and the field is invariant with respect to changes in the brane position, it is said to be brane invariant. The case of gauge fields is more interesting because the fundamental fields have unphysical components related to the gauge ambiguity. This leads to an interplay between gauge transformation and brane transformation which is the manifestation of the well known concepts of small and large gauge transformations. These concepts become very clearly stated in the language of multivalued fields and branes. 

The elements presented in this work thus lead to a natural description of defects. We discuss the well known result that the insertion of defects in the system amounts to a deformation of the original action with the introduction of terms representing singularities in the domain of the fundamental fields. This result, cast in the language of multivalued fields and branes, naturally furnishes the correct formulation. In that way we were able to reproduce the results of \cite{Marino:1980rn} for the computation of correlation functions of disorder operators. Also the concepts of Wilson and t' Hooft operators in abelian theories can be naturally cast in this language. 

The interplay of gauge symmetry and brane symmetry is important for the study of superconductors and its ensuing vortices. We showed that open and closed vortices can be introduced and its corresponding correlation functions computed.  In fact, for the case of topological superconductors, we saw that in the description of the electromagnetic response the so called chiral vortices are an important aspect of the phenomenology. These vortices can be naturally incorporated taking into account the multivaluedness of the axion-like field in the effective theory introduced in \cite{Qi:2012cs}. 

All the results presented here are valid for the case where the multivalued character of the field can be expressed as a direct sum of a regular part and a singular part. For the case of gauge theories it is thus suitable for application to abelian systems only. The definition of the covariant derivative also shows that the introduction of matter is straightforward for bosons and fermions. But it would be interesting to explore further the introduction of fermions due to the matrix nature of the "polar decomposition" of a fermionic field. The generalization of our prescription to non-abelian systems is less straightforward because of the mixing between the singular and regular parts and the interpretation of such terms becomes less clear.

\section*{Acknowledgements}
The Conselho Nacional de Desenvolvimento Científico e Tecnológico (CNPq-Brazil), the Faperj, Fundação de Amparo à Pesquisa do Estado do Rio de Janeiro, the SR2-UERJ and the Coordenação de Aperfeiçoamento de Pessoal de Nível Superior (CAPES) are gratefully acknowledged for financial support. M. S. Guimaraes is supported by the Jovem Cientista do Nosso Estado program - FAPERJ E-26/202.844/2015, is a level PQ-2 researcher under the program Produtividade em Pesquisa-CNPq, 307801/2017-9 and is a Procientista under SR2-UERJ.

\end{document}